\documentclass[prc,aps,twocolumn,nofootinbib]{revtex4}
\usepackage[utf8]{inputenc}
\usepackage{xcolor}
\usepackage{physics}
\usepackage{graphicx}
\usepackage{hyperref}
\usepackage{qcircuit}
\usepackage{float} 
\usepackage{amsmath,amssymb,amsthm}

\renewcommand{\expval}[1]{\langle #1 \rangle}

\begin{document}
\bibliographystyle{ieeetr}

\title{Solving the Lipkin model using quantum computers with two qubits only with a hybrid quantum-classical technique based on the Generator Coordinate Method}

\author{Yann Beaujeault-Taudi\`ere}
\email{beaujeaul@ijclab.in2p3.fr}
\affiliation{Universit\'e Paris-Saclay, CNRS/IN2P3, IJCLab, 91405 Orsay, France, \\ Laboratoire Leprince-Ringuet (LLR), \'Ecole polytechnique, CNRS/
 IN2P3, F-91128 Palaiseau, France
}

\author{Denis Lacroix } \email{lacroix@ijclab.in2p3.fr}
\affiliation{Universit\'e Paris-Saclay, CNRS/IN2P3, IJCLab, 91405 Orsay, France}

\begin{abstract}
The possibility of using the generator coordinate method (GCM) using hybrid quantum-classical algorithms with 
reduced quantum resources is discussed. The task of preparing the basis states and calculating the various kernels involved in the GCM is assigned to the quantum computer, while the remaining tasks, such as finding the eigenvalues of a many-body problem, are delegated to classical computers for post-processing the generated kernels. This strategy reduces the quantum resources required to treat a quantum many-body problem.  
We apply the method to the Lipkin model. Using the permutation symmetry of the Hamiltonian, we show that, ultimately, only two qubits is enough to solve the problem regardless of the particle number. 
The classical computing post-processing leading to the full energy spectrum can be made using standard generalized eigenvalues techniques by diagonalizing the so-called Hill-Wheeler equation. 
As an alternative to this technique, we also explored how the quantum state deflation method can be adapted 
to the GCM problem. In this method, variational principles are iteratively designed 
to access the different excited states with increasing energies.  
The methodology proposed here is successfully applied to the Lipkin model with a minimal size of two qubits for 
the quantum register. The performances of the two classical post-processing approaches with respect to the 
statistical noise induced by the finite number of measurements and quantum devices noise are analyzed. Very satisfactory results for the full energy spectra are obtained once noise corrections techniques are employed.  
\end{abstract}
\date{\today}
\maketitle

\section{Introduction}

In the contemporary landscape of computational science, quantum computing has emerged as a burgeoning field characterized by its potential to address problems that elude classical computational methodologies. 
Many-body interacting systems are particularly adapted for testing quantum computers \cite{McC16,Cao19,McA20,Bau21,Bha21,End21} due to the fact that quantum computers themselves are often built from quantum many-body interacting systems \cite{Ayr23}.   

Presently, intensive international efforts are concentrated on two pivotal fronts: firstly, the development of quantum algorithms tailored to solve specific many-body problems, and secondly, the validation of these algorithms, most often on rather schematic models. To quote some of the test Hamiltonian that are relevant in the context of nuclear physics and for which quantum algorithms have been developed, we mention the Lipkin-Meshkov-Glick (LMG) Hamiltonian (sometimes referred to as the Lipkin model) \cite{Cer21,Rob21a,Rom22,Hla22,Hla23,Rob23}, and its extension known as the Agassi model \cite{Per22}, or the pairing/Richardson Hamiltonian \cite{Lac20,Kha21,Rui21,Rui22,Rui23,Lac23}.    
These models, while based on simplifying assumptions compared to the complexity of the nuclear many-body problem, serve as invaluable testbeds due to their ease of implementation on classical computers. Over the years, scientists have honed their physical intuition by exploring these models, yet they occasionally encompass complexities that prove challenging to translate into quantum computational frameworks. In parallel 
to the advances in quantum computation, these models are also shedding illuminating insights from a quantum information perspective \cite{Fab21,Fab22,Mom23}, which could also be interesting in the more general context of the entanglement properties in nuclear physics \cite{Rob21,Joh23,Per23,Gu23}.

In this study, we explore the possibility of using the Generator Coordinate Method (GCM) in the context of quantum computing. The GCM approach is nowadays widely used on classical computers, with some successes in the nuclear physics context \cite{Rin80,Bla86,Ben03,Rob18,She19}. The various types of applications of the GCM theory, ranging from obtaining ground or excited state properties or the possibility to describe the dynamical path towards fission, vividly showcase the versatility and great potential inherent to this approach. As far as we know, the possible interest of using a GCM-based approach on quantum computers 
has only been discussed very recently in Ref. \cite{Zhe23}, where some advantages of using the GCM-based approach, including the reduction in the circuit depth compared to currently used techniques, have been pointed out. 

We further discuss here the possibility of developing a GCM-inspired hybrid quantum-classical algorithm to study the energy spectrum 
of a many-body Hamiltonian. The GCM approach leads to a generalized eigenvalue problem formulated in a non-orthogonal set of many-body states, i.e., the so-called Hill-Wheeler equation \cite{Hil53}.  
A natural strategy for sharing the tasks between classical and quantum processor units (called hereafter CPU and QPU, respectively), is to allocate solely to the quantum computer the estimation of the different matrix elements entering the Hill-Wheeler equation, while
eigenvalues are obtained on the classical computer using the QPU inputs. Our focus narrows down to the Lipkin model, a recent cornerstone in benchmarking quantum algorithms, particularly within the domain of nuclear physics. We show that due to the permutation invariance of this model Hamiltonian \cite{Lip65,Mes65,Gli65}, together with the use 
of the Generalized Wick Theorem \cite{Bla86}, all kernels 
can be estimated with a minimal number of 2 qubits, a significant reduction compared to other quantum algorithms 
proposed so far. We analyze two possible techniques to perform the classical post-processing. Specifically, we 
either use the standard diagonalization technique or the deflation approach \cite{Hig19} that does not rely 
on any diagonalization. Both techniques have advantages and possible caveats when used on 
quantities estimated from quantum measurements. The efficiency of the two post-processing 
with respect to statistical errors are analyzed in detail.  

Our preliminary results on the Lipkin model framework exhibit promising outcomes, underscoring the potential efficacy of this approach in the quantum computational domain.

\section{Generator Coordinate method applied on quantum computers}
\label{sec:gcm}

\subsection{Brief introduction to the GCM approach with generator coordinate}
\label{sec:GCMstandard}

The GCM approach is rather well documented \cite{Rin80,Ben03}, and we only give here the key ingredients.  
This method consists in approximating eigenstates of a many-body problem written in the form: 
\begin{eqnarray}
| \Psi \rangle & = & \int_{\mathbf q} f({\mathbf q}) | \Phi ({\mathbf q}) \rangle d {\mathbf q} ,\label{eq:gcmansatz}
\end{eqnarray}
where $| \Phi ({\mathbf q}) \rangle$ is a set of non-orthogonal many-body states depending on a few collective coordinates $\{ {\mathbf q} = (q_1, \cdots , q_c)\}$. Different techniques can be used to generate 
the subsets $\{| \Phi ({\mathbf q}) \rangle\}$. These states will be referred to as generating states below. In some cases, like in the context of symmetry restoration, the states are generated through a pre-defined Thouless transformation, and the mixing parameters $f({\mathbf q})$
are fixed \cite{Bla86,Rin80}. In this case, the integral form (\ref{eq:gcmansatz}) corresponds to a projector to a good symmetry sector. The advantage of using such a form for post-processing to enforce symmetries 
in quantum computing has already been pointed out in several works \cite{Tsu20,Kha21,Sek22,Tsu22,Lac23,Rui23b}. 
In some other cases, relevant, for instance in the exploration of phase coexistence or large amplitude collective adiabatic motion, the states are generated using constrained mean-field methods, where a potential energy landscape is built up by imposing a 
set of constraints $\{q_i\}_{i=1,c}$ on selected one-body operators $\{ \hat Q_i \}_{i=1,c}$, i.e., such that 
$q_i = \langle \Phi ({\mathbf q})| \hat Q_i |  \Phi ({\mathbf q}) \rangle $. While other options are rarely explored, we see that the GCM approach offers enormous flexibility in choosing the subsets of states. 
For eigenvalue problems, the GCM equations to be solved are obtained using the Rayleigh-Ritz variational 
principle \cite{Bla86}, and can be written generically as:
\begin{eqnarray}
    \int_{{\mathbf q}'} d {{\mathbf q}'} 
    \left[ {\cal H}({{\mathbf q}}, {{\mathbf q}'}) - E {\cal N}({{\mathbf q}}, {{\mathbf q}'})\right] f({{\mathbf q}'}) = 0, \label{eq:gcmeigen}
\end{eqnarray}
where ${\cal H}({{\mathbf q}}, {{\mathbf q}'})$ and ${\cal N}({{\mathbf q}}, {{\mathbf q}'})$ are the Hamiltonian and norm kernels respectively, defined as:
\begin{eqnarray}
    \left\{ 
    \begin{array}{l}
     {\cal H}({{\mathbf q}}, {{\mathbf q}'}) =  \langle \Phi ({\mathbf q}) |\hat H| \Phi({\mathbf q}') \rangle  \\
     \\
     {\cal N}({{\mathbf q}}, {{\mathbf q}'}) =  \langle \Phi({\mathbf q}) | \Phi({\mathbf q}') \rangle   
    \end{array}
    \right. .\label{eq:kernels}
\end{eqnarray}
The appearance of the norm kernels is because the generated states $| \Phi( {\mathbf q} ) \rangle$ are not orthogonal from each other. The solution of the generalized eigenvalue equation, known as the Hill-Wheeler  equation, is usually achieved  
using a finite number of generated states $N_{\rm st}$. Such a solution is standardly made today on a classical computer and consists in (i) diagonalizing the norm matrix $\mathcal{N}_{{\mathbf q}, {\mathbf q}'} \equiv {\cal N}({{\mathbf q}}, {{\mathbf q}'})$. This step gives ultimately a new set of orthogonal states associated with a set of norm eigenvalues $\{ \xi_n \}_{n=1,N_{\rm st}}$;  
(ii) part of the eigenstates having their eigenvalues lower than a certain threshold $\varepsilon$ are not further considered in the following steps to remove the eventual redundancy
contained in the non-orthogonal basis; and (iii) solve the eigenvalue problems in this new, possibly truncated, basis, to obtain approximate eigenvalues and eigenstates by diagonalizing the Hamiltonian in this basis. Since this approach relies heavily on diagonalization solvers, we will refer to it as GCM-Diag below.     

\subsection{Hybrid quantum-classical algorithm for the GCM}

One of the advantages of the GCM approach is to use a subset of states, with several states $N_{\rm st}$, 
that is usually much smaller than the total Hilbert space size necessary to consider for solving exactly the problem. The quality of the approximate energy spectra obtained heavily depends on  
the proper optimization of the states used in the decomposition (\ref{eq:gcmansatz}). Provided that $N_{\rm st}$
remains small enough, and that kernels are estimated with sufficient precision, Eq. (\ref{eq:gcmeigen}) can be solved without
specific difficulty on a classical computer. In most, if not all, applications to nuclear physics of the GCM today, the generated states are restricted to quasi-particle vacua. One reason is that the corresponding mean-field energies, i.e., the diagonal kernels
can be obtained using the Wick theorem in terms of the generalized density matrix. This greatly facilitates the minimization of these energies under the generator coordinates constraints through, for instance,  
efficient gradient descent methods. 
A second reason is that for two different quasi-particle vacua, thanks to the generalized Wick Theorem \cite{Bal69}, 
the off-diagonal elements of the kernels (\ref{eq:kernels}) can also be estimated without too much difficulty. This simplification stems from the specific Lie group algebra inherent
to transforming one state into another (see appendix E of Ref. \cite{Rin80}). 

These simplifications in estimating the kernels break down when the generating states $| \Psi({\mathbf q}) \rangle$ are not anymore 
independent quasi-particle states and, as far as we know, this option has not been explored in classical computers, 
probably due to the increase in complexity. Quantum computers offer the perspective to (i) explore a wide class of quantum ans\"atz and (ii) within the context of noisy quantum computers, great progress is being made in optimizing these ans\"atze,
using, for instance, Variational Quantum Eigensolver (VQE)-based techniques \cite{Per14} (see also Ref. \cite{Ayr23} and references therein). Having the long-term goal to apply the GCM approach using generated states beyond the independent 
quasi-particle picture, the following hybrid quantum-classical methods can be proposed:
\begin{enumerate}
    \item A circuit that depends on a set of parameters $\boldsymbol\theta= \{(\theta_1 , \cdots, \theta_M \}$ and associated to a unitary transformation $U(\boldsymbol\theta)$ is first 
    selected to prepare the different states that are used later for the configuration mixing in Eq. (\ref{eq:gcmeigen}), 
    where now $\boldsymbol\theta$ replaces the collective coordinates ${\mathbf q}$, i.e., we now have the generating states defined as:
    \begin{eqnarray}
        | \Phi (\boldsymbol\theta) \rangle &\equiv& U(\boldsymbol\theta) \bigotimes_{j=0}^{q-1} | 0_j \rangle 
    \end{eqnarray}
    on a certain number of qubits $q$. Here $| 0_j \rangle $ and $|1_j \rangle$
  denote the two states associated the $j^{th}$ qubit. The entanglement properties and expressivity of the ans\"atz will 
  depend on the choice of the $U(\boldsymbol\theta)$.
 \item Eventually, as in the usual GCM, the energy is minimized under a set of constraints. This can be done independently for each state 
    by using the VQE approach with the cost function:
    \begin{eqnarray}
        {\cal C}(\boldsymbol\theta) =& \langle H \rangle_{\boldsymbol\theta} - \sum_{k=1}^{c} \xi_k \left(\langle Q_k  \rangle_{\boldsymbol\theta}  - q_k \right), 
    \end{eqnarray}
    where the compact notation $\langle . \rangle_{\boldsymbol\theta} \equiv \langle \Phi (\boldsymbol\theta) |.| 
    \Phi (\boldsymbol\theta) \rangle$. The $\xi_k$ plays the role of the Lagrange multipliers that enforce the different constraints. By modifying the values for the set of parameters $\{q_k \}_{k=1,c}$, one can explore different regions of the collective energy landscape and generate a subset of states 
    $\{ | \Phi (\boldsymbol\theta _\alpha)\rangle \}_{\alpha=1,N_{\rm st}}$.  
    \item For each couple $(\boldsymbol\theta _\alpha , \boldsymbol\theta _\beta)_{\alpha,\beta=1,N_{\rm st}}$, 
    a new circuit is used employing standard quantum algorithms to estimate the two kernels 
    ${\cal N}(\boldsymbol\theta _\alpha , \boldsymbol\theta _\beta)$
     and ${\cal H}(\boldsymbol\theta _\alpha , \boldsymbol\theta _\beta)$. Further discussions on this aspect are given 
     in section \ref{sec:lipkin}. 
    \item Finally, once the $N_{\rm st}(N_{\rm st}+1)$ matrix elements have been obtained, the Hamiltonian and overlap matrices 
    are treated on a classical computer to deduce an approximate energy spectrum and eventually associated eigenvectors. For this task, the diagonalization procedure discussed previously can be used. An alternative post-processing will be introduced in section \ref{sec:def}.  
\end{enumerate}

As a first pilot application, we show below how this strategy can be implemented in the specific case of 
the Lipkin model. In this context, we still stick to the traditional GCM implementation where the generated states are simple Slater determinants because, in this model, there is no need to use a more complex quantum ansatz, even to solve the problem exactly. We note in passing that no specific advantage is expected of using quantum computers with respect to solving this model case fully on a classical computer. Nevertheless, such early-stage applications are very useful both to identify possible future caveats when using quantum computers for configuration mixing and to illustrate that the GCM-guided methods can 
outperform recent approaches recently proposed to solve the Lipkin model, especially regarding the required 
quantum resources.  

\section{Application to the Lipkin model}
\label{sec:lipkin}

\subsection{Basic aspects of the LMG model and its qubit encoding}

As an illustration of the hybrid protocol discussed in the previous section, 
we test the method in the LMG model \cite{Lip65}. This model consists of a 
set of $N$ two-level systems. 
Let us introduce the notation $\{ | 0_\alpha \rangle , |1_\alpha \rangle$\} respectively with energies $\{ -\varepsilon/2, \varepsilon/2 \}$ for the two 
single-particle states in a given $2$-level system, with $\alpha=1,N$. Introducing the associated 
creation operator $a^{\dagger}_{0/1, \alpha}$, the LMG Hamiltonian reads:    
\begin{eqnarray}
H &=& \varepsilon J_z + \frac{V}{2} (J_+ J_+ + J_- J_-) \equiv H_{\rm 1b} + H_{\rm 2b},
\label{eq:hamillipkin} 
\end{eqnarray}
with the definition of the total spin operators $(J_x, J_y, J_z)$ [together with $J_{\pm} = J_x \pm i J_y$]:
\begin{eqnarray}
    J_z &=& \frac{1}{2} \sum_{\alpha} \left( a^\dagger_{1, \alpha} a_{1, \alpha} - a^\dagger_{0, \alpha} a_{0, \alpha}\right), \nonumber \\
    J_{+} &=& \sum_{\alpha} a^\dagger_{1, \alpha} a_{0, \alpha}, ~~J_{-} = J^\dagger_+ . \nonumber
\end{eqnarray}
We assume here that the number of particles is equal to the number of $2$-level systems, in which case, each 
2-level system has exactly one particle. An important symmetry of the problem is its invariance with respect to swapping any two indices $(\alpha, \beta)$ labeling two different 2-levels. This invariance, referred to as permutation invariance below, implies that the many-body eigenstates are also eigenstates of the total spin $\mathbf{J}^2$. Thanks to this property, the many-body problem can be solved on a classical computer
in a reduced basis of $(N+1)$ states $|J, M \rangle$ formed by the common eigenstates of $\mathbf{J}^2$ and 
$J_z$. Ultimately, a $(N+1) \times (N+1)$ matrix should be diagnonalized. The numerical effort can eventually be reduced by noting that the odd- and even-parity blocks form two independent sub-blocks \cite{Lip65,Mes65,Gli65}.  

In pioneering attempts to solve the LMG problem on a digital quantum computer, three types of encoding 
have been proposed: (i) the brute force Jordan-Wigner  \cite{Jor28,Lie61} transformation (JWT) \cite{Rom22}; (ii) the direct SU(2) encoding of each 2-level \cite{Cer21}; and (iii) the compact encoding based on first quantization, where the $M$ values label the $|J,M \rangle$ states \cite{Rob21a}. The different encodings differ significantly in terms of the 
resulting Hamiltonian in the computational basis, leading respectively to (i) $q=2N$, (ii) $q=N$, and (iii) 
$q=\lceil \log_2 N \rceil$ \cite{Hla22,Hla23}. The last encoding, which is the only one using the 
permutation invariance of the problem, evidently outperforms the other ones in terms of quantum register 
size. Compared to other encodings, it also leads to compact unitary (tridiagonal) matrices that might lead to a significant number of gates and/or, eventually the need to add ancillary qubits to implement them. 

Here, we follow the encoding (ii) and will show that one can still take advantage of the 
permutation invariance in our scheme, even if it is not enforced in the first place. We
define the Pauli operators associated with the 2-level $\alpha$ as:
\begin{eqnarray}
\left\{
\begin{array}{l}
    X_\alpha =   | 1_\alpha \rangle \langle 0_\alpha| + | 0_\alpha \rangle \langle 1_\alpha|  ,  \\
    \\
Y_\alpha = i | 1_\alpha \rangle \langle 0_\alpha|  - i | 0_\alpha \rangle \langle 1_\alpha|, \\
\\
Z_\alpha = | 0_\alpha \rangle \langle 0_\alpha| - | 1_\alpha \rangle \langle 1_\alpha|
\end{array}
\right. \label{eq:paulisu2}
\end{eqnarray}
and complement it by the qubit identity $I_\alpha$. With this, the total spin components $(J_x, J_y, J_z)$ simply 
identify with the sum of the corresponding Pauli matrices divided by $2$. Using the natural SU(2) encoding, the LMG Hamiltonian writes in terms of Pauli strings as 
\begin{eqnarray}
    H = \frac{\varepsilon}{2}\sum^{N}_{\alpha=1} Z_\alpha + \frac{V}{4} \sum^{N}_{{\alpha \neq \beta}} (X_\alpha X_{\beta} - Y_\alpha Y_{\beta}), \label{eq:LMGPauli}
\end{eqnarray}
and should a priori be solved in a quantum register of size $q=N$ qubits. We show below that the GCM-based approach solves this problem with much fewer qubits in practice.  

\subsection{Coherent generating states}

We know from classical computing application of the GCM \cite{Hol73,Rin80,Rob92,Sev06} that a natural choice for describing models like the LMG model is the SU(2) coherent states \cite{Zha90}. These states form an overcomplete basis for this problem, and using them as generating states is expected to converge to the exact result when sufficient states are considered. Coherent states are usually defined using two angles.  In the present work, we consider a simplified version of these states using a single angle, assuming that 
generating states are a function of a single collective coordinate $\theta$ and are defined as:
\begin{eqnarray}
    | \Phi(\theta) \rangle &=& e^{i\theta J_y} \bigotimes_{\alpha=0}^{q-1} | 0_\alpha \rangle \equiv U(\theta) | {\mathbf 0} \rangle,  \label{eq:gentheta}
\end{eqnarray}
where we introduce the notation $| {\mathbf 0 } \rangle = \bigotimes_{\alpha=0}^{q-1} | 0_\alpha \rangle  $. This SU(2)
$\rightarrow$ SO(2) reduction stems from the fact that the eigenvectors of Hermitian operators can be taken real. Hence, there is no need to introduce a complex phase.
States generated in this way remain simple states that could be written as tensor products with:
\begin{eqnarray}
    | \Phi(\theta) \rangle &=&  \bigotimes_{\alpha=0}^{q-1} \left[ R^\alpha_Y(\theta) | 0_\alpha \rangle \right],  \label{eq:tensortheta}
\end{eqnarray}
where $R^{\alpha}_Y(\theta)$ denotes the standard one qubit Y-rotation acting on the $\alpha$ qubit. We also see from this expression that each coherent state is individually permutation invariant. Note also, that the state (\ref{eq:tensortheta}) corresponds in the many-body context to Slater determinants $|\Psi(\theta)\rangle = \prod_\alpha c^\dagger_\alpha(\theta) |-\rangle$, where $|-\rangle$ is the Fock space particle vacuum and 
$c^\dagger_\alpha(\theta)$ is obtained from the original $(a^\dagger_{0,\alpha}, a
^\dagger_{1,\alpha})$ by a simple $2 \times 2$ Bogolyubov transformation (not shown here).  

For the LMG case, using the one-parameter dependent generating states as discussed here, there is no need to optimize the states before the GCM-guided procedure. Here, we directly consider a grid in $\theta$ space. Specifically, we consider a set of $L$ equally spaced values of $\theta$, denoted 
by $\theta_l = -\pi (1-\delta(L)) + l \Delta \theta$ with $\Delta \theta = 2\pi (1-\delta(L)) / (L-1)$. The factor $\delta(L)=1/L$ is a simple prescription to avoid $\theta_0=\theta_{L-1}$, which would make the corresponding trial states identical. Even though the states given by Eq. (\ref{eq:gentheta}) do not have a well-defined parity, using a regular grid centered on $\theta=0$ makes it possible to respect this symmetry at the level of many-body observables through the linear combinations.
\begin{eqnarray}
    | \Phi(\theta) \rangle_\pm = \frac{| \Phi(\theta) \rangle \pm | \Phi(-\theta) \rangle}{\sqrt{2}}.
\end{eqnarray}

In the following, we will simply use the notation $| \Phi_l \rangle = | \Phi(\theta_l) \rangle \equiv U_l | {\mathbf 0} \rangle$, and write the GCM ans\"atz (\ref{eq:gcmansatz}) assuming a discretized form as:
\begin{eqnarray}
| \Psi \rangle &=& \sum_{l=0}^{L-1} f_l | \Phi_l \rangle. \label{eq:startinggcm}
\end{eqnarray} 
In the specific LMG model, we know that the ans\"atz above can describe the exact solution 
provided that $L \ge  (N+1)$. 

\subsection{Reducing the quantum resources for kernels using permutation symmetry}

Let us consider a general observable $O$, that we write as a linear combination of unitaries 
such that $O= \sum_k g_k V_k$, where $V_k$ is a product of Pauli matrices acting on the 
set of qubits. The expectation value of $O$ with the ans\"atz (\ref{eq:startinggcm}) is given by:
\begin{eqnarray}
    \langle \Psi | O | \Psi \rangle &=& \sum_{l,l',k} f^*_l f_{l'} g_k \langle \Phi_l |V_k| \Phi_{l'} \rangle .
\end{eqnarray}
Using the strategy employed in the VQE, each matrix element $\langle \Phi_l |V_k| \Phi_{l'} \rangle$
can be estimated using a separate circuit. Taking advantage of the relation
\begin{eqnarray}
\langle \Phi_l |V_k| \Phi_{l'} \rangle &=& \langle {\mathbf 0} | U^\dagger_l V_k U_{l'} |  {\mathbf 0}  \rangle \equiv \langle V_k\rangle_{ll'}, 
\end{eqnarray} 
the real and imaginary parts of these quantities can be obtained using standard Hadamard tests \cite{Nie00} 
at the price of adding a single ancillary qubit. 
Implementing the operation $U_l$ given by Eq. (\ref{eq:gentheta}) is rather straightforward for the Lipkin model since it consists in $q$ independent Y-rotation of each qubit. For a more general situation, discussions on the cost and practical aspects to prepare Slater determinants 
or more generally of quasi-particle states can be found in \cite{Aru20} and \cite{Dal18}. 

The $(q+1)$ register needed to encode the LMG problem in the SU(2) encoding, although it scales linearly with the number $N$ of 2-levels, still
prevents from using the SU(2) for large $N$ values on quantum computers. 
In \cite{Cer21}, only applications up to $N=2$ have been possible on real 
devices. Even today, using the most compact encoding \cite{Hla22,Hla23} has allowed to perform applications up to $N=8$ using $q=3$ qubits. 

In the quest to apply the GCM-guided method, we realized that the explicit encoding of the problem within the SU(2) scheme on $(q+1)$ 
qubits for the LMG is not necessary. Indeed, starting from the expression of the Hamiltonian (\ref{eq:LMGPauli}), and 
using the permutation invariance of each individual generating state, 
the quantities $\langle I_\alpha \rangle_{ll'}, \langle Z_\alpha \rangle_{ll'}$, $\langle X_\alpha X_\beta \rangle_{ll'}$
or $\langle Y_\alpha Y_\beta \rangle_{ll'}$ are independent on the choice of the qubit $\alpha$ or on the pairs of qubits 
$(\alpha, \beta)$ respectively. This reduces the number of observable expectation values to only $4$ for each pair of states $(l,l')$. Then, using the simple tensorial nature of the generating state, one can ultimately recast the Hamiltonian 
expectation value as:
\begin{eqnarray}
\label{eq:Hkernelone}
 \langle H \rangle_{ll'} &=& \frac{\varepsilon N}{2} i^{N-2}_{ll'} \left[ i_{ll'} z_{ll'}  + \frac{\chi}{2} \left( x^2_{ll'} - y^2_{ll'}\right)  \right],
\end{eqnarray}
while for the norm kernel, we simply have 
\begin{eqnarray}
\langle I \rangle_{ll'} = i^{N}_{ll'}. \label{eq:Normkernelone}
\end{eqnarray}
In the Hamiltonian expectation value, we used the parameter $\chi = V(N-1)/\varepsilon$, usually introduced in the mean-field approximation of the LMG model.  The different quantities $p_{ll'}$ with  $p=(i,x, y, z)$ are kernels obtained from 
2 states that are expressed on a single qubit register. Specifically, we have for these states $| \phi_l \rangle = e^{-i\theta_l Y/2} | 0 \rangle \equiv u(\theta_l) | 0 \rangle$ and: 
\begin{eqnarray}
    p_{ll'} &=& \langle 0 | u^\dagger(\theta_l) P u(\theta_{l'}) | 0 \rangle, \label{eq:kernelone}
\end{eqnarray}
with $P= (I,X,Y,Z)$. The real and imaginary parts of the different one-body kernels $p_{ll'}$ can be 
obtained using the Hadamard tests shown in Fig. \ref{fig:Hadamard_tests} that only requires $2$ qubits 
and with only 3 CNOT gates. 

In summary, 
we have just shown that the use of the GCM on a quantum computer, together with employing the permutation invariance of the problem and the fact that the generating states are simple tensor product states, leads to a tremendous reduction 
of the quantum resources needed to express the GCM kernels, independently of the number $N$ 
of 2-level considered. This holds even if we use a fermion-to-qubit encoding that 
does not explicitly use the permutation invariance symmetry. Note that the compact encoding of Ref. \cite{Hla22,Hla23} explicitly taking advantage of this symmetry to reduce the quantum resources 
requires more qubits. In this case, the qubit number scales as $ \lfloor \log_2 (N+1) \rfloor$. This number 
is reduced to $q=2$ here. 

\subsection{Estimation of the kernels with two qubits}
\label{sec:hadamard}

The circuits from which the real and imaginary parts of $p_{ll'}$, based on the Hadamard test, are given in Fig. \ref{fig:Hadamard_tests}. 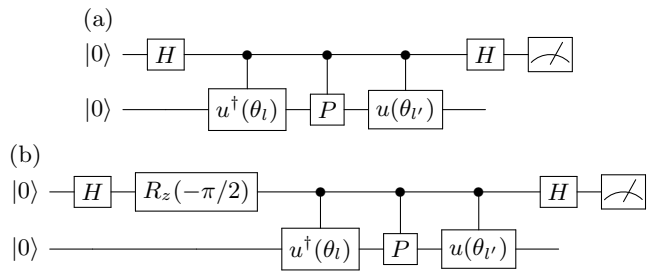
\begin{figure}[H]
    \centering
    \[
    \Qcircuit @C=1em @R=.7em {
        \mbox{(a)} \\
        \ket{0} & & \gate{H} & \ctrl{1} & \ctrl{1} & \ctrl{1} & \gate{H} & \meter \\
        \ket{0} & & \qw & \gate{u^\dagger(\theta_l)} & \gate{P} & \gate{u(\theta_{l'})} & \qw & }
        \]   
    \[
    \Qcircuit @C=1em @R=.7em {
        \mbox{(b)} \\
        \ket{0} & & \gate{H} & \gate{R_z(-\pi/2)} & \ctrl{1} & \ctrl{1} & \ctrl{1} & \gate{H} & \meter \\
        \ket{0} & & \qw & \qw & \gate{u^\dagger(\theta_l)} & \gate{P} & \gate{u(\theta_{l'})} & \qw & }
        \]   
    \caption{Quantum circuits using the Hadamard (a) and modified Hadamard (b) tests, respectively, employed to extract the real and the imaginary part of the expectation values of $p_{ll'}$, with $p \in (i,x,y,z)$ associated to the operator 
    $P=(I,X,Y,Z)$. The real and imaginary parts are recovered 
    from the difference of probabilities $p_0$ and $p_1$ to measure $0$ or $1$, respectively in the ancillary 
    qubit. }
    \label{fig:Hadamard_tests}
\end{figure}

\begin{figure}[H]
    \centering
    \includegraphics[width=\linewidth]{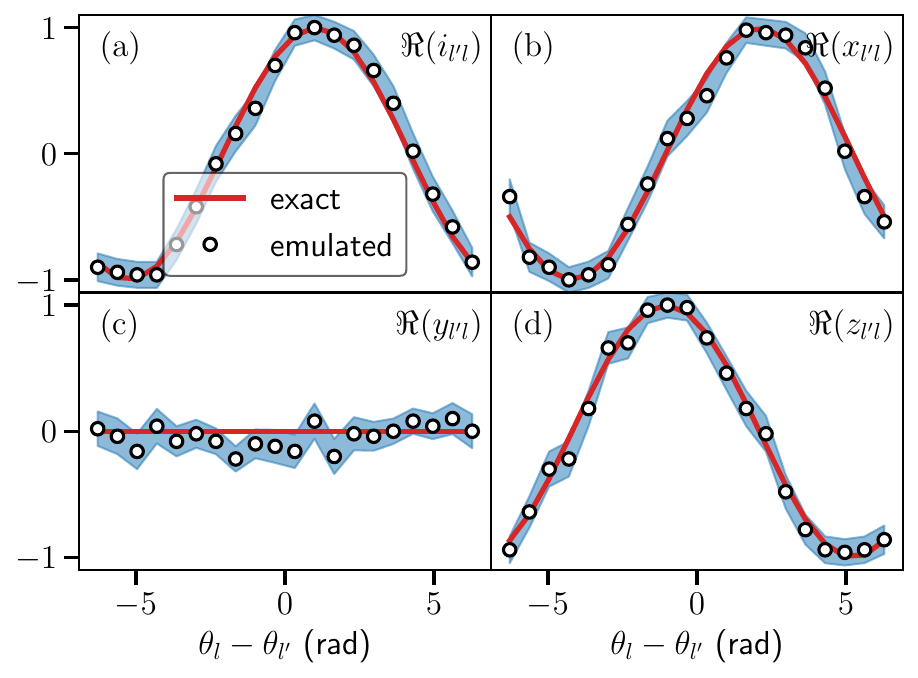}
    \caption{Real parts of the quantities (a) $i_{l'l}$ ,  (b) $x_{l'l}$, (c) $y_{l'l}$, and (d) $z_{l'l}$ as a function 
    of $\theta_l$ for $\theta_{l'}=\pi/3$. In each panel, the exact values are reported with red solid lines, while the average values obtained from 
    $N_{\rm sh} = 100$ shots measurements are shown with open black circles. The shaded blue area represents the standard deviation that is proportional to $1/{\sqrt{N_{\rm sh}}}$. The results have been obtained by simulating a noiseless fault-tolerant quantum computer using the IBM-Qiskit package \cite{Qis21}.
    }
    \label{fig:kernelone}
\end{figure}

We show in Fig. \ref{fig:kernelone}, illustrations of the real parts of the matrix elements $i_{ll'}$, $x_{ll'}$, $y_{ll'}$
and $z_{ll'}$ obtained with the circuit of Fig. \ref{fig:Hadamard_tests} using a finite number of shots $N_{\rm sh}$. 
Similar estimates can be obtained for the imaginary parts (not shown).   
In the following, the different shots are labelled by the integer $\lambda= 1, N_{\rm sh}$ and we denote generically the specific value of one of the kernels $p^{(\lambda)}_{ll'}$ for the event $\lambda$. 
The reported values correspond to the average over the total number of shots, defined as (for compactness, 
we simply use below the notation $p^{(\lambda)}_{ll'} = p^{(\lambda)}$):
\begin{eqnarray}
\overline{p^{(\lambda)}} &=& \frac{1}{N_{\rm sh}} \sum_{\lambda}  p^{(\lambda)}. \label{eq:globalmean}
\end{eqnarray}
These averages are subject to a global standard deviation scaling as $1/\sqrt{N_{\rm sh}}$.
We also report in the figure (shaded area) the usual standard deviation that is defined as:
\begin{eqnarray}
    \sigma_p(N_{\rm sh}) &=& \frac{1}{\sqrt{N_{\rm sh}}} \sqrt{\sum_{\lambda}  \left[ p^{(\lambda)}\right]^2 - \overline{p^{(\lambda)}}^2}.
\end{eqnarray} 

We see in Fig. \ref{fig:kernelone} that even a small number of shots 
gives a rather good estimate of the different one-qubit kernels with limited deviations. 

\subsection{Application of the standard GCM diagonalization solver}

Once the different one qubit kernels reported in Fig. \ref{fig:kernelone} are obtained for each couple of angles $(\theta_l, \theta_{l'})$, 
one can deduce the norm and Hamiltonian many-body kernels from Eqs. (\ref{eq:Normkernelone}) and (\ref{eq:Hkernelone}), respectively, for any number $N$ of
2-levels. Note, however, that an error on the one-qubit kernel will induce an increasing error on the many-body kernels as $N$ 
increases. In addition, the size of the Hilbert space relevant for the permutation invariance is equal to $(N+1)$ and, therefore, 
changes with $N$. Accordingly, for a fixed number of generating states $L$, with the equidistant prescription, we expect to be able 
to solve exactly the problem for $N$ such that $L \ge (N+1)$. If $L < (N+1)$, only approximate solutions will be obtained that will depend on the 
retained set of angles $\{\theta_l \}_{l=0,L-1}$. 

\begin{figure}[htbp]
    \centering
    \includegraphics[width=\linewidth]{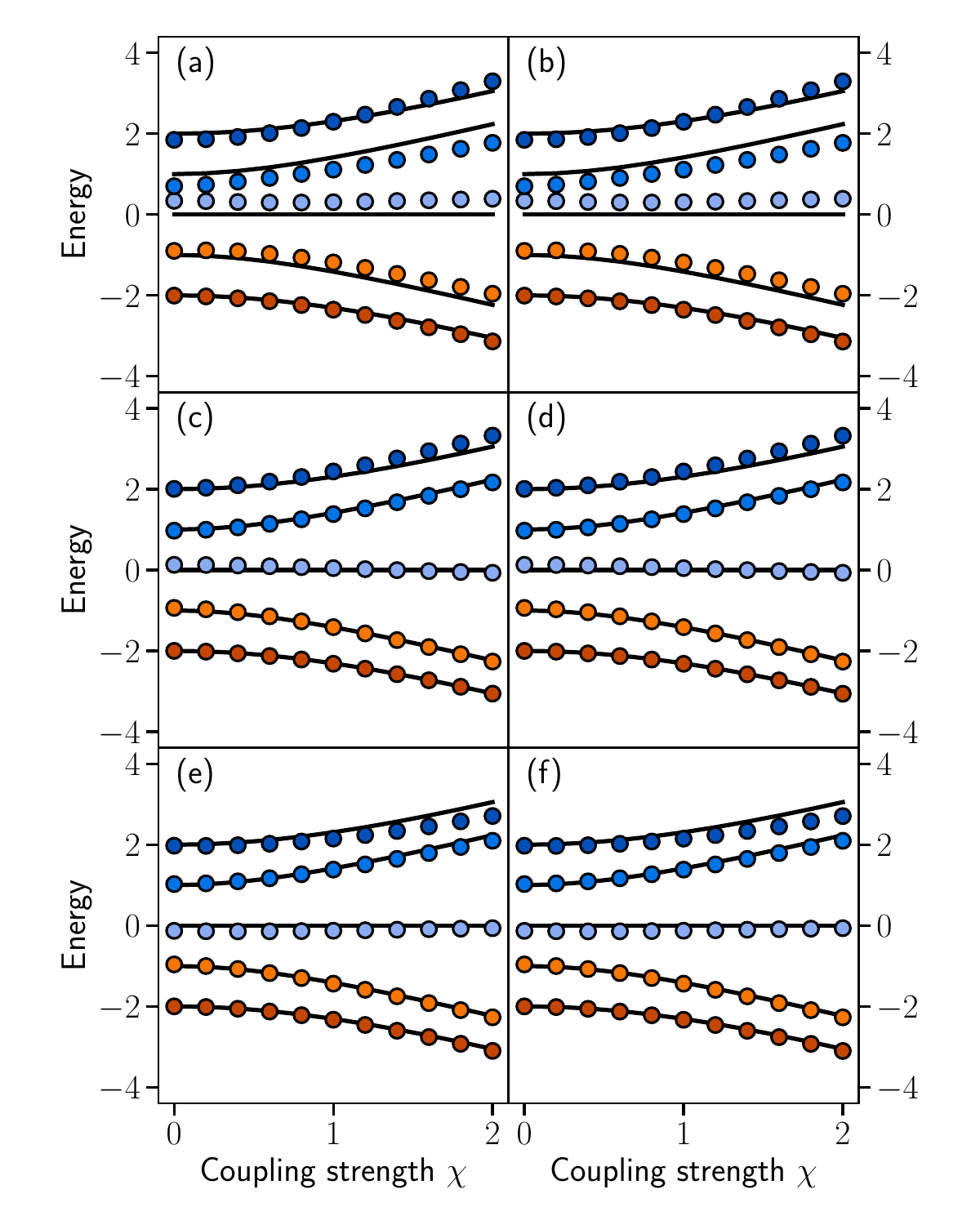}
    \caption{Exact (solid lines) versus approximate energies (symbols) for the LMG model as a function of $\chi$ for the case 
    $N=4$. In panel (a), (c), and (e) are shown the results obtained using the hybrid method with the 
    GCM-Diag technique using $10^2$, $10^3$, and $10^4$ shots in the estimation of each 
    one-body kernels. In panels (b), (d), and (b) are the results obtained from the same set of
    shots but using the deflation method instead of the diagonalization.}
    \label{fig:gcmdiagvsgcmVQDN4}
\end{figure}  

\begin{figure}[htbp]
    \centering
    \includegraphics[width=\linewidth]{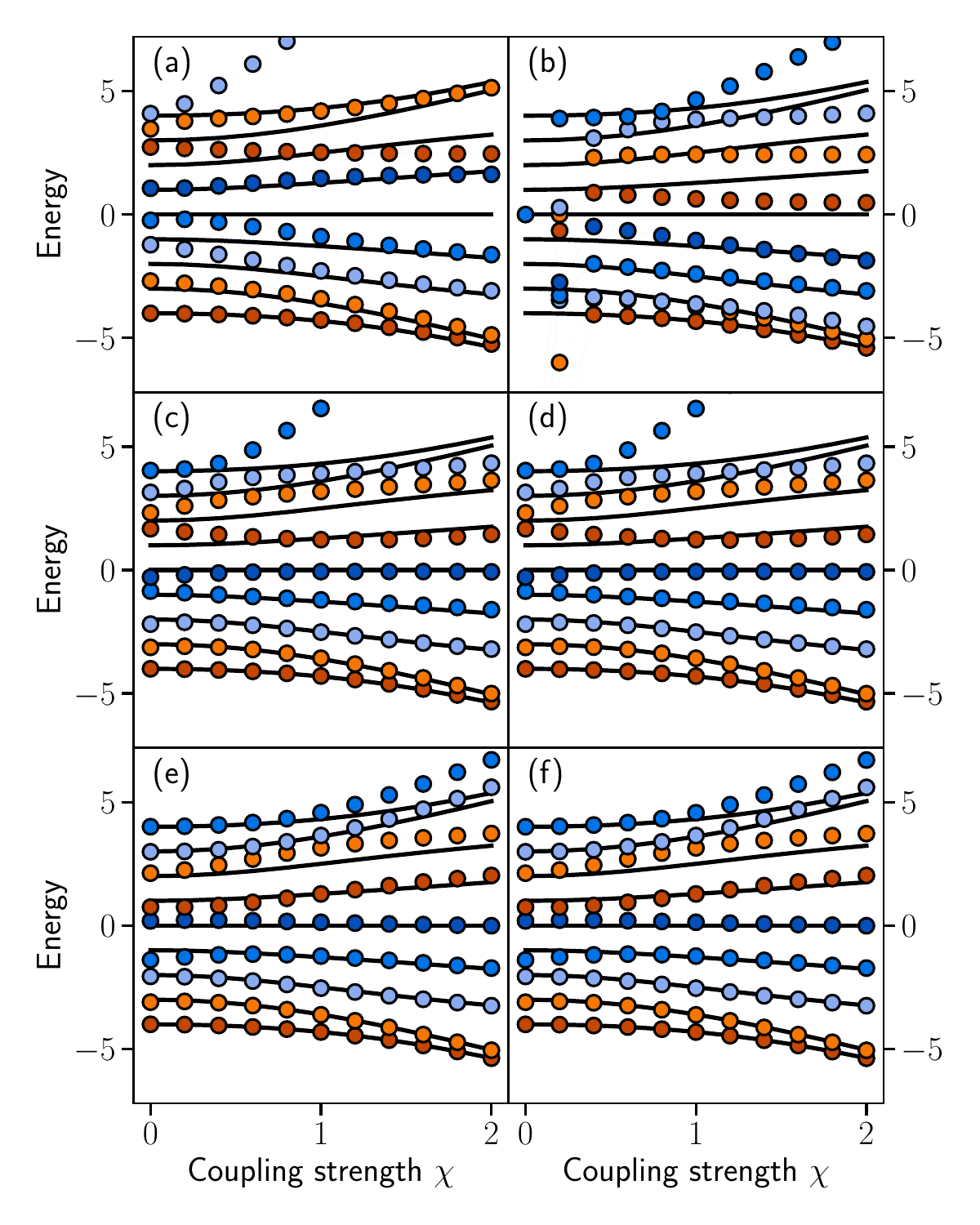}
    \caption{Same as Fig. \ref{fig:gcmdiagvsgcmVQDN4} for the $N=8$ case. 
    Note that, in this case, a better precision on the one-body kernels is necessary
    compared to the $N=4$ case to achieve a similar reproduction of the eigenvalues. For this 
    reason, slightly higher numbers of shots are used to obtain the figure: $10^4$, $10^5$, and $10^{6}$ shots for panels (a-b), (c-d) and (e-f) respectively.}
    \label{fig:gcmdiagvsgcmVQDN8}
\end{figure}  

The GCM-Diag method, which is the one standardly used on classical computers, consists in solving the Hill-Wheeler equation (\ref{eq:gcmeigen}) that takes the discretized form here, for all $l$:
\begin{eqnarray}
 \sum_{l'} \mathcal{H}_{ll'} f_{l'} &=&  E \sum_{l'} f_{l'} \mathcal{N}_{ll'} .
\end{eqnarray}
Today, such an equation is always solved on a classical computer using a brute-force diagonalization procedure that consists of first diagonalizing $\mathcal{N}$ to transform the problem into a standard eigenvalue problem in a new orthonormal basis and then diagonalizing the Hamiltonian in this basis. This method is well documented and has been discussed, for instance, in the quantum computing context (see, for instance, Refs. \cite{Sta20, Bha21a,Bha21b,Rui22}) as a practical way to solve iterative many-body problems using a quantum-classical technique based on a quantum state expansion picture.  

We show in panels (a), (c), and (e) of Fig. \ref{fig:gcmdiagvsgcmVQDN4} and 
\ref{fig:gcmdiagvsgcmVQDN8}, some results obtained from the brute-force diagonalization technique using 
increasing numbers of shots from top to bottom panels.  
The values $L=5$ and $L=9$ are used respectively for $N=4$ and $N=8$, 
such that in the limit of an infinite number of events, we expect to get all eigenstates of the problem. 
We see that, provided that enough shots are used to get one-body kernels with good precision, the results of the GCM-Diag reproduce the exact eigenvalues of the Hamiltonian. From this study, we see that the hybrid strategy based on the estimate of many-body kernels using solely two qubits, followed by classical post-processing where many-body kernels are then reconstructed and then 
a brute force GCM scheme is used, is validated. 

Still, from our extensive studies during this work, we have observed that very small errors on the one-body kernels 
lead to large errors in the many-body kernels that increase with $N$. 
This is illustrated by the fact that a higher number of shots, i.e., higher precision 
on the one-body kernels, was necessary to obtain results shown in \ref{fig:gcmdiagvsgcmVQDN8} 
to achieve a similar precision on the eigenenergies compared to Fig. \ref{fig:gcmdiagvsgcmVQDN4}.

One critical aspect that is also the source of difficulty in standard classical computing simulation of GCM stems from the inversion of the overlap matrix. Proper information of the information content of the non-orthogonal basis 
relies on the use of a threshold $\epsilon$ for its eigenvalue, denoted previously as $\{\xi_n \}_{n=1,N_{\rm st}}$. 
A typical threshold used in general is of the order of $10^{-3}-10^{-5}$. We anticipate that a precision on the eigenvalues at this level 
will be extremely difficult to achieve in noisy quantum computers. 

To avoid the necessity to invert the norm kernel, we also explored 
the quantum deflation technique that gives access to approximate eigenstates without the norm matrix inversion. This technique is discussed below. 

\subsection{Combining the GCM and quantum state deflation}
\label{sec:def}

As an alternative to the GCM diagonalization technique, we have explored the possibility of performing the classical 
post-processing using the deflation technique of Ref. \cite{Hig19}, that (i) we slightly adapt here to the specific case where the 
trial state vectors are written as a function of a subset of non-orthogonal states, and (ii) we pay specific attention to the fact that we should avoid 
the inversion of the overlap matrix. 

The Variational Quantum Deflation (VQD) technique is an iterative procedure to generate eigenstates with increasing values of energies. The GCM problem is considered 
in this work, and this technique is applied as follows:
\begin{enumerate}
    \item The quantum state deflation is initiated using a VQE technique using the trial state vector given by Eq. (\ref{eq:startinggcm}). Specifically, we use the following cost function that is minimized: 
    \begin{eqnarray}
        {\cal C}_0 &=& \langle \Psi |H| \Psi \rangle - \lambda_0  \left[ \langle \Psi | \Psi \rangle - 1 \right]^2
    \end{eqnarray}
    where $\lambda_0$ is a Lagrange multiplier used to constrain the trial state norm to be one. In this expression, the 
    expectation values of the energy and norms are computed from the many-body kernels that are themselves deduced from the one-body kernels
    obtain from the quantum simulation.
    After this step, we end up with an approximate ground state denoted by $| \Psi(0) \rangle$ and associated with the 
 mixing coefficients $\{ f_l(0)\}_{l=0,L-1}$. 
    \item Approximate eigenstates with increasing energies are then built using the following iterative scheme. Let us assume that we already obtained a set of approximations for the $k=0,$ $K-1$ lowest eigenstates, denoted by $\{| \Psi(k) \rangle \}$, each associated to a set of mixing coefficients $\{ f_l(k) \}_{l=0, L-1}$. The next state $| \Psi(K+1) \rangle$ is obtain by minimizing the cost function:
    \begin{eqnarray}
        {\cal C}_{K}  &=& \langle \Psi (K)   | H | \Psi (K) \rangle  \nonumber \\
        &-&
\lambda_K  \left[ \langle \Psi (K)  | \Psi (K)  ) \rangle - 1\right]^2 \nonumber \\
  &-& \sum_{j=0,K-1}  \beta_K (j) | \langle \Psi (j) | \Psi (K) \rangle | ^2 
\end{eqnarray}    
where we again recognize the Lagrange multiplier added to normalize the state, while, in addition, a set of $j=0,$ $K-1$ 
new Lagrange multipliers is added to force the constraint $\langle \Psi (j) | \Psi (K) \rangle  = 0$ for $j < K$. Again, the cost function 
${\cal C}_K$ is minimized assuming the trial ans\"atz of Eq. (\ref{eq:startinggcm}), leading to the set of optimal mixing coefficients $\{ f_l (K) \}_{l=0, L-1}$. The procedure is then iterated until the desired number of states $N_{\rm st}$ is reached. Note that in practice 
$N_{\rm st}$ cannot be higher than the rank of the overlap matrix itself.      
\end{enumerate}
The above scheme is referred to as GCM-VQD below.  

We show in panels (b), (d), and (f) of Figs. \ref{fig:gcmdiagvsgcmVQDN4} and \ref{fig:gcmdiagvsgcmVQDN8}
results obtained with the deflation method as an alternative to the diagonalization solver. Note that 
the same one-body kernels are used in both cases. It is worth mentioning that since the same generator states and the same Hamiltonian matrix elements are used, we expect that both methods strictly lead 
to the same results within the accuracy of the approach. 
One difference between the deflation approach and the diagonalization is that the minimization of the variational principle is made through a specific optimizer. Here, we used the limited-memory Broyden–Fletcher–Goldfarb–Shanno optimizer \cite{Byr95, Zhu97}, implemented in the SciPy library. To avoid divergences in the optimization process, it is necessary to bind the value of the mixing coefficients $f_l$. In practice, we constrain them to be within the bounds $-2\leq f_l \leq 2$, as we found this to be more numerically stable than the restrictions $\abs{f_l}\leq 1$ due to the use of a non-orthogonal basis set. We also used the COBYLA method, with, in general, slightly inferior results. As shown in \cite{Hig19}, the Lagrange multipliers $\{\beta_K(j)\}$ do not need to be adjusted through the procedure, as long as an estimate for the energy difference between eigenstates is available. Here, we use $\beta_K(j)=10N$, which fulfills that condition for all the studied values of the coupling constants. Another possible issue with the deflation method is that eventual errors of low-lying states automatically impact 
higher energy states due to the iterative variational principle strategy. We observed in practice 
that the VQD approach might have difficulties converging, especially in the weak coupling regime, 
if a small number of shots is used for the one-body kernels. 
Nevertheless, provided that one-body kernels are obtained with sufficient precision, we see that  
the results converge towards exact results in a similar way as the GCM-Diag case. 
More generally, the precision of the two approaches on the eigenvalues are generally similar. 
\begin{figure}[htbp]
    \centering
    \includegraphics[width=0.8\linewidth]{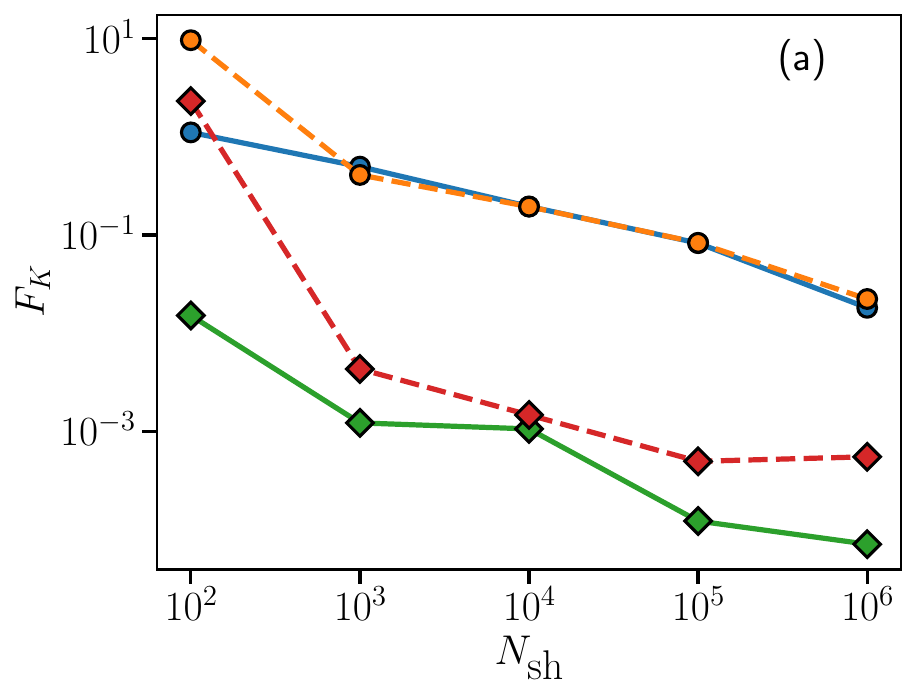} \\
    \includegraphics[width=0.8\linewidth]{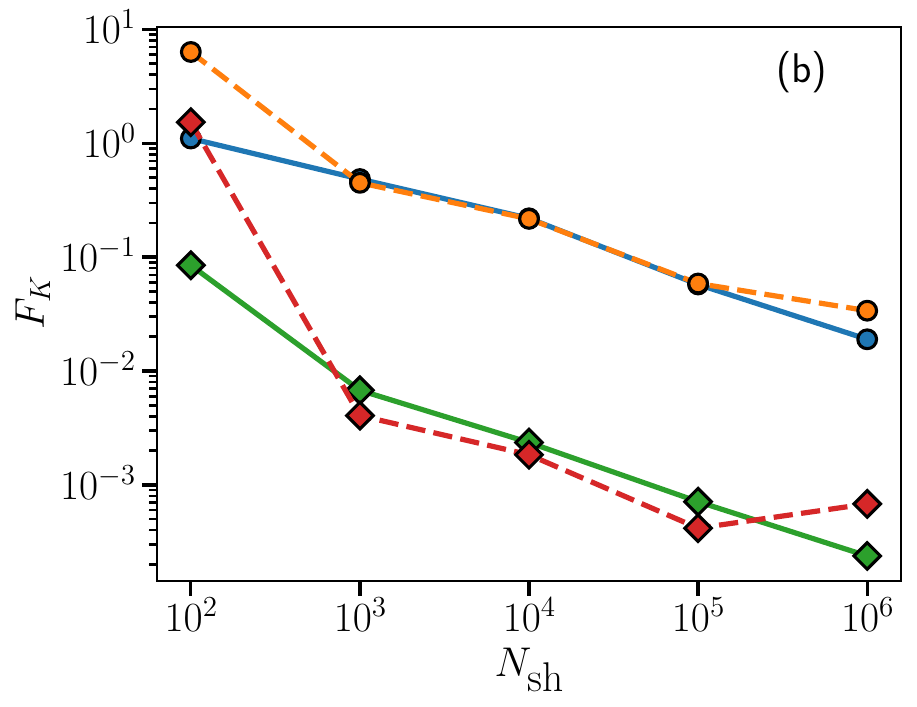} \\
    \includegraphics[width=0.8\linewidth]{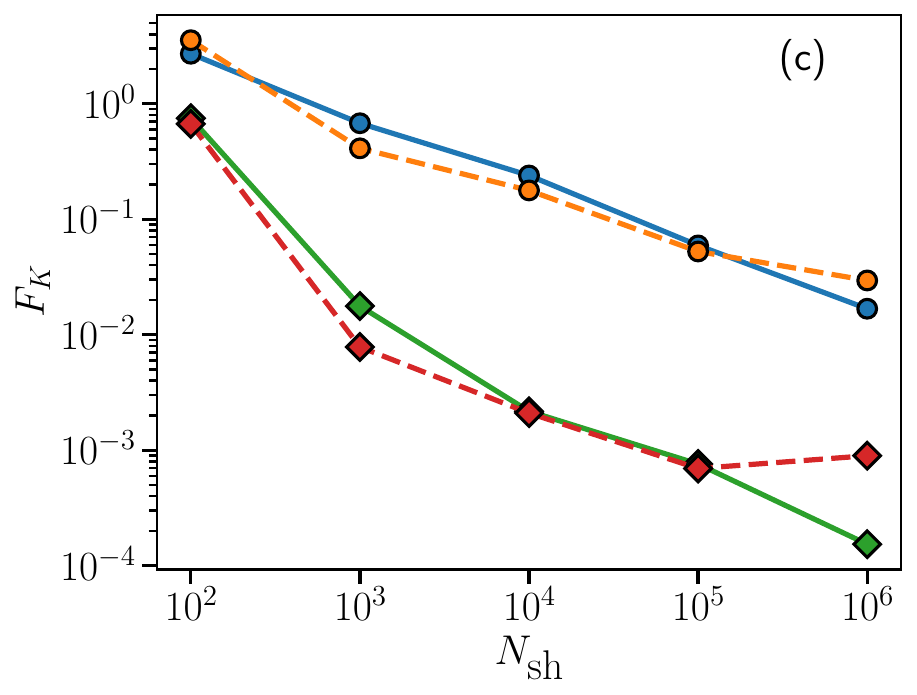} 
    \caption{$F_K$ given by Eq. (\ref{eq:endev}) obtained by GCM-Diag (solid lines) and GCM-VQD 
    (dashed lines) for $N=8$ and for (a) $\chi=0.2$, (b) $\chi=1$, and (c) $\chi=2$. The round and square symbols correspond to $K=N+1$ (full spectrum) and $K=1$ (ground state only), respectively.
    }
    \label{fig:devenergy}
\end{figure}  
This is illustrated in Fig. \ref{fig:devenergy} where we show the quantity 
\begin{eqnarray}
    F_K &=& \frac{1}{\abs{E_\text{gs}} K} \sqrt{\sum_{\alpha=1}^{K} \left( \overline{E_\alpha} - E_\alpha\right)^2}, \label{eq:endev}
\end{eqnarray}
where $\{ \overline{E_\alpha}\}$ are the set of approximate energies obtained either by diagonalization 
or state deflation while  $\{E_{\alpha} \}$ denote the exact energies. $E_\text{gs}$ denotes the ground state energy.  
We see in Fig. \ref{fig:devenergy}, that the diagonalization gives slightly better results also 
compared to the QSD when a small number of shots are used. However, it should be kept in mind that the 
use of $10^2$ shots leads to very bad results in both cases. 

\section{Tests on noisy quantum computer emulators}

In this section, we test the quantum GCM approach resistance to the quantum device noise. To this end, we use the {\it Lagos} backend emulator 
provided by the IBM Quantum platform \cite{Qis21}. The key parameters of the backend are given on Table \ref{tab:lagos}. 

\begin{table}[H]
    \centering
    \begin{tabular}{||c|c|c|c||}
        \hline
        $T_1$  ($\mu s$) &  $T_2$ ($\mu s$) & Readout error & CNOT error\\
        \hline
        84.23 & 28.45 & 1.44$\times 10^{-2}$ & 8.79$\times 10^{-3}$\\
        \hline
    \end{tabular}
    \caption{Mean values of the main physical parameters of the Lagos backend.}
    \label{tab:lagos}
\end{table}

\begin{figure}[H]
    \centering
    \includegraphics[width=\linewidth]{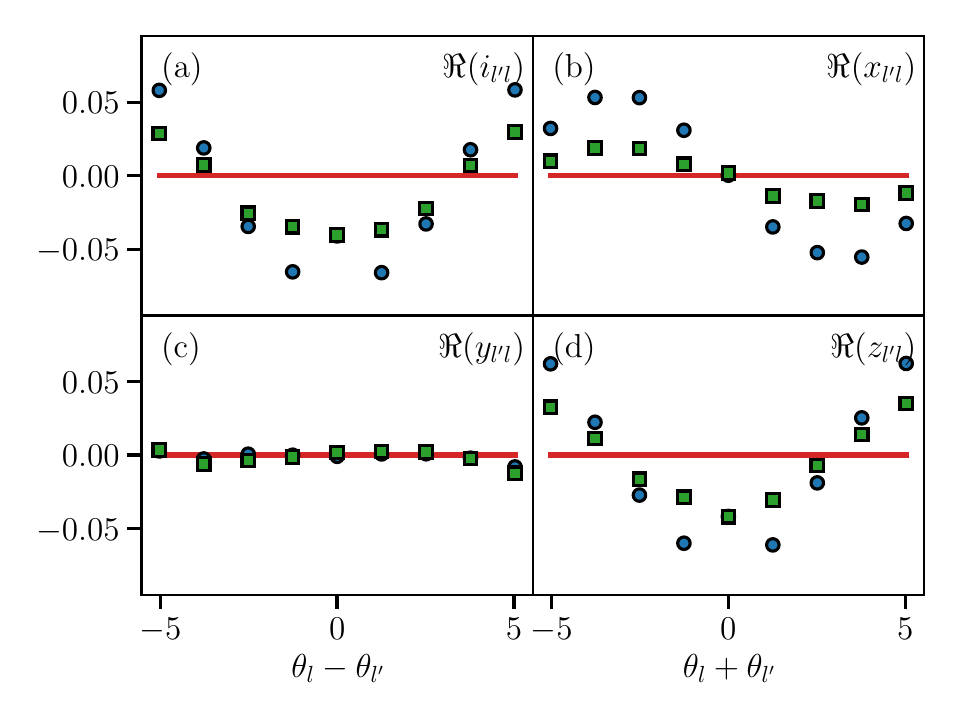}
    \caption{Error in the real parts of the quantities (a) $i_{l'l}$,  (b) $x_{l'l}$, (c) $y_{l'l}$, and (d) $z_{l'l}$ as a function 
    of $\theta_l\pm\theta_{l'}$, without error mitigation (blue circles) and after zero-noise extrapolation and scaling (green squares). Each point represents the average from $N_\text{sh}=10^5$ measurements. The horizontal red lines indicate the zero error limit. }
    \label{fig:noise_mitigation}
\end{figure} 

\begin{figure}[H]
    \centering
    \includegraphics[width=\linewidth]{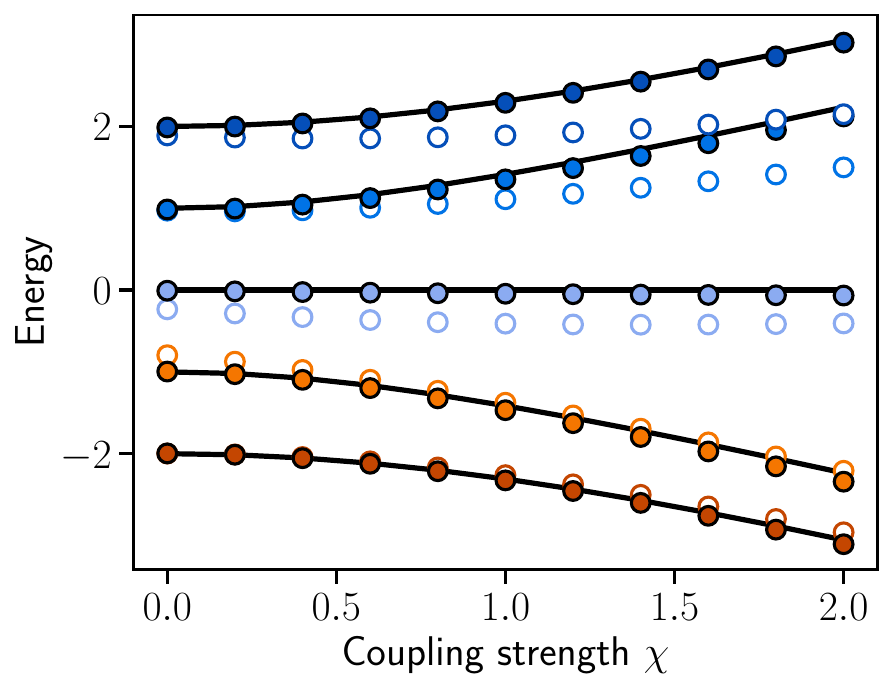}
    \caption{Energies obtained for the LMG model with $N=4$ as a function of the coupling strength $\chi$ with the quantum GCM-Diag approach using the kernels deduced from noisy simulations without (open circles) and with (filled circles) error corrections. The error corrections contain here both the ZNE and scaling approach discussed in the text. The solid lines correspond to the exact energies. Each one-body kernel is computed using $10^5$ measurements.}
    \label{fig:energyN4noise}
\end{figure}

\begin{figure}[H]
    \centering
    \includegraphics[width=\linewidth]{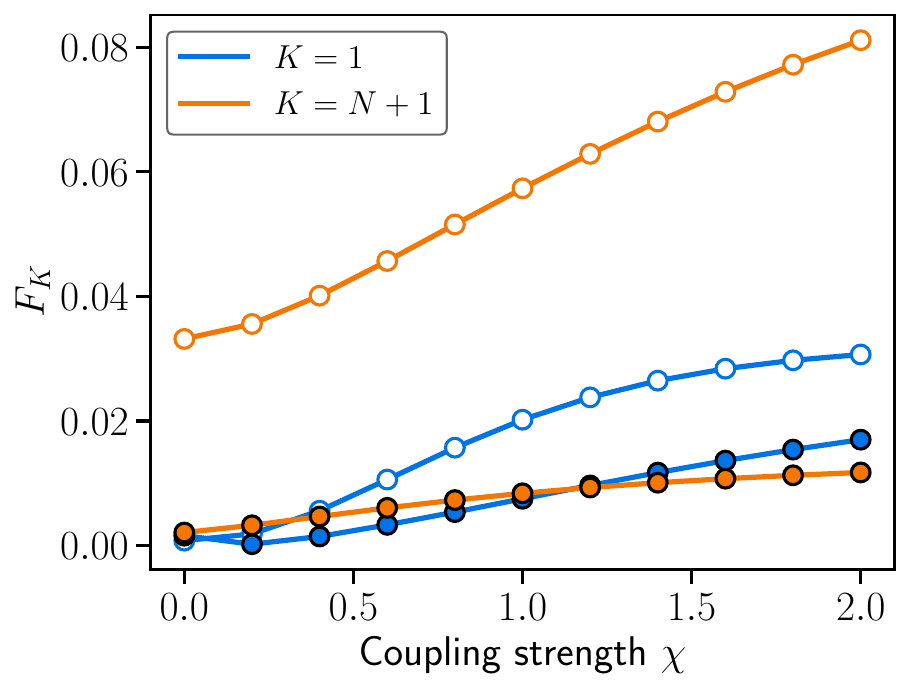}
    \caption{$F_K$ for the ground state energy (blue) and for the full spectrum (orange). The open circles points correspond to results without error corrections, and the filled one to the results with ZNE+scaling
    corrections. All points correspond to the GCM-Diag. Application of the GCM-VQD (not shown) method leads to results that are equivalent to the diagonalisation approach. }
    \label{fig:lagoserrorN4}
\end{figure}

We show in Fig. \ref{fig:noise_mitigation} the deviation of the one-body kernels obtained from the quantum simulations 
compared to the exact results (solid lines in Fig. \ref{fig:kernelone}). Except for the $y_{ll'}$ kernels, for which the noisy results 
rather well match the exact case, we see some deviations for the other kernels that could be of the order of $6-7 \%$ of the expected values.  
Note that Fig. \ref{fig:noise_mitigation} has been obtained using $10^5$ shots for each points (compared to $100$ shots in Fig. \ref{fig:kernelone}), ensuring negligible statistical 
errors compared to the errors induced by the quantum device's imperfections. 
An error of a few percent on the one-body kernels, since it propagates in the many-body kernels, has a
significant impact on the results obtained using the quantum GCM calculation. Correcting the error even by a factor of two can reduce the error on the many-body kernels by almost one order of magnitude for $N=4$. This is illustrated in Fig. \ref{fig:energyN4noise}. In this figure, we see that even with the presence of noise, the ground state and first excited states are rather close to the exact energies. However, the reproduction of the excited state's energies rapidly degrades with increasing excitation energy, whatever is the coupling strength. The corrections on the one-body kernels strongly improve the results, in particular for the excited states. The quantitative absolute deviation of the calculated energies for the ground state and for all states is shown in Fig. 
\ref{fig:lagoserrorN4}. From this, the quantum GCM approach is quite efficient for low-lying states, and can also reproduce the low-lying states with good precision, provided the error one one-body kernels is small enough. For the ground state and after noise mitigation, an error of less than  $2\%$ for all 2-body coupling strength is observed. 

To complement the present application in the presence of noise, we incorporated some standard noise mitigation techniques. 
Specifically, we used the standard zero-noise extrapolation technique (ZNE) \cite{Li17,Tem17}, which we implement by repeating the circuits \ref{fig:Hadamard_tests} and their inverse $k=0,1,2$ times. In the GCM application, the circuit depth is much smaller than the error rate of the different gates, and the total execution time of the circuit remains much smaller than the mean relaxation ($T_1 = 84.23$ $\mu s$) and dephasing ($T_2 = 28.45$ $\mu s$) times. 
The circuit error is thus essentially proportional to the total depth, and we extrapolate to zero noise using simply a linear fit. 

In addition to the ZNE technique, a second technique is employed to correct the error, which will be referred to below as the scaling method. In the LMG model, the task of the quantum computer is minimal and reduces to perform simple unary rotation for which we know some specific properties should hold. Among 
them, we should have the exact equalities $\expval{I}=\expval{Z}=1$ for $\theta_l=\theta_l'=0$, and $\expval{X}=i\expval{Y}=1$ for $\theta_l=-\theta_l'=\pi/2$. Due to noise, these equalities are only approximately fulfilled. A simple method to improve the 
calculation is to re-scale the noisy results such that these constraints are exactly fulfilled.  
Corrected one-body kernels are reported in Fig. \ref{fig:noise_mitigation}, while eigenstates energies 
after the ZNE+scaling procedure are shown as open circles in Fig. \ref{fig:energyN4noise}. Finally, the absolute error on the low-lying state after noise correction is shown in Fig. \ref{fig:lagoserrorN4}.  
The ZNE+scaling correction procedure systematically reduces the error by about a factor of two for the one-body kernels, and ultimately strongly improves the energies of the many-body eigenstates.

\section{Conclusion}

Guided by the GCM approach standardly used in nuclear physics in classical computing, 
a hybrid quantum-classical method is proposed to obtain the ground state and excited state
of a many-body Hamiltonian. In our strategy, the only task attributed to the quantum computer 
is the computation of the Hamiltonian and norm many-body kernels, while the eigenenergies are deduced from them 
by classical post-processing. The strategy is illustrated in the LMG model case, which has recently become
a milestone for quantum computing in nuclear physics. Two techniques are used for the post-processing: 
the direct diagonalization and the deflation method. Both methods give comparable results that match the 
exact solution, provided that enough generator states are used and that the kernels are estimated with enough accuracy. The present work demonstrates that the GCM-guided approach works properly. Most importantly, 
we illustrate with the LMG model that, using the inherent symmetries of the Hamiltonian to estimate the 
kernels, the proposed strategy can lead to a significant reduction of the quantum resources. Specifically, 
two qubits are needed together with a very small number of operations, outperforming all 
methods that have been proposed so far to solve the LMG model. 

In the LMG model, besides the proof of principle that the approach works, we do not expect a specific 
advantage in using quantum computers compared to a fully classical computing treatment. Still, we have learned 
from this simple example that, using the GCM technique, one can greatly reduce the quantum resources needed to estimate the kernels 
by using the symmetry of the problem. The LMG model 
is also simple enough that we do not need to optimize the generator states. In a more complex situation, 
the full strategy will generate the states through an optimization process, such as using 
the VQE approach. Such VQE based on parameterized states has become one of the most efficient tools 
on noisy quantum computers. 

We also stress that the GCM approach, being independent of the specific class of ans\"atz used to prepare the trial states, can be used very straightforwardly on top of any single-reference quantum algorithm at a moderate increase of the circuit complexity.

Finally, we would like to mention the great versatility of the GCM-based technique. In the present work, 
we used the standard strategy to perform GCM that is based on using simple quasi-particle states. One 
motivation to use these states stems from the possibility of taking advantage of the Generalized Wick Theorem, which greatly simplifies the computation of kernels on a classical computer. Using more complex
states that incorporate more correlations significantly increases the cost of estimating the kernels
on classical computers. We anticipate that quantum computers will allow estimating kernels for wider classes 
of many-body states and extend the current applicability of GCM approaches.

\section{Acknowledgments }

We thank G. Hupin, E.A. Ruiz Guzman, and J. Zhang for the discussions during this work. 
This project has received financial support from the CNRS through the AIQI-IN2P3 project and the P2IO LabEx (grant ANR-10-LABX-0038).
This work is part of 
HQI initiative (www.hqi.fr) and is supported by France 2030 under the French 
National Research Agency award number ``ANR-22-PNQC-0002''.
We acknowledge the use of IBM Q cloud as well as the use of the Qiskit software package
\cite{Qis21} for performing the quantum simulations.


\begin{thebibliography}{99}


\bibitem{McC16} J.R. McClean, J. Romero, R. Babbush, A. Aspuru-Guzik, {\it The theory of variational hybrid quantum-classical algorithms}, New J. Phys. {\bf 18}, 023023 (2016). 


\bibitem{Cao19} Y. Cao, J. Romero, J.P. Olson, M. Degroote, P.D. Johnson, M. Kieferov\'a, I.D. Kivlichan, T. Menke, B. Peropadre, N.P.D. Sawaya, S. Sim, L. Veis, A. Aspuru-Guzik, {\it Quantum chemistry in the age of quantum computing}, Chem. Rev. {\bf 119}, 10856 (2019). 

\bibitem{McA20} S.McArdle, S.Endo, A.Aspuru-Guzik, S.C.Benjamin, X.Yuan, {\it Quantum computational chemistry},  Rev. Mod. Phys. {\bf 92}, 015003 (2020). 


\bibitem{Bau21} B.Bauer, S.Bravyi, M.Motta, G.Kin-LicChan, {\it Quantum algorithms for quantum chemistry and quantum Quantum Materials Science},  Chem. Rev. {\bf 120}, 12685 (2020). 


\bibitem{Bha21} K. Bharti, A. Cervera-Lierta, T.H. Kyaw, T. Haug, S. Alperin-Lea, A. Anand, M. Degroote, H. Heimonen, J.S. Kottmann, T. Menke, W.-K. Mok, S. Sim, L.-C. Kwek, A. Aspuru-Guzik, {\it Noisy intermediate-scale quantum algorithms}, Rev. Mod. Phys. {\bf 94}, 015004 (2022). 

\bibitem{End21} S. Endo, Z. Cai, S.C. Benjamin, X. Yuan, {\it Hybrid quantum-classical algorithms and quantum error mitigation},  J. Phys. Soc. Jpn. {\bf 90}, 032001 (2021). 

\bibitem{Ayr23} T. Ayral, P. Besserve, D. Lacroix and A. Ruiz Guzman, 
{\it Quantum computing with and for many-body physics}, Eur. Phys. J. {\bf A 59} (2023). 



\bibitem{Cer21} Michael J. Cervia, A. B. Balantekin, S. N. Coppersmith, Calvin W. Johnson, Peter J. Love, C. Poole, K. Robbins, and M. Saffman, {\it Lipkin model on a quantum computer}
Phys. Rev. {\bf C 104}, 024305 (2021) 

\bibitem{Rob21a} Robbins, Kenneth and Love, Peter J., {\it Benchmarking near-term quantum devices with the variational quantum eigensolver and the Lipkin-Meshkov-Glick model}, Phys. Rev. {\bf A 104}, 022412 (2021)

\bibitem{Rom22} A. M. Romero, J. Engel, Ho Lun Tang, Sophia E. Economou, {\it Solving Nuclear Structure Problems with the Adaptive Variational Quantum Algorithm}, Phys. Rev. {\bf C 105}, 064317 (2022).  


\bibitem{Hla22} M.Q. Hlatshwayo, Y. Zhang, H. Wibowo, R. LaRose, D. Lacroix, E. Litvinova, 
{\it Simulating excited states of the Lipkin model on a quantum computer}, Phys. Rev. {\bf C 106}, 024319 (2022).

\bibitem{Hla23} Manqoba Q. Hlatshwayo, John Novak, Elena Litvinova, {\it Quantum benefit of the 
quantum equation of motion for the strongly coupled many-body problem}, arXiv:2309.10179 



\bibitem{Rob23} C. E. P. Robin, M. J. Savage, {\it Quantum simulations in effective model spaces (I): 
Hamiltonian Learning-VQE using digital 
quantum computers and application to the 
Lipkin-Meshkov-Glick model},  Phys. Rev. {\bf C 108}, 024313 (2023) 


\bibitem{Per22} P. P\'erez-Fern\'andez, J.-M. Arias, J.-E. Garc\'ia-Ramos, L. Lamata, {\it A digital quantum simulation of the Agassi model}, Phys. Lett. {\bf B 829}, 137133 (2022).


\bibitem{Lac20} D. Lacroix, 
{\it Symmetry-Assisted Preparation of Entangled Many-Body States on a Quantum Computer}, Phys. Rev. Lett. {\bf 125}, 230502 (2020).

\bibitem{Kha21} A. Khamoshi, T. Henderson, and G. Scuseria, 
{\it Correlating AGP on a quantum computer}, Quant. Sci. Technol. {\bf 6}, 
014004 (2021).

\bibitem{Rui21} E. A. Ruiz Guzman and D. Lacroix, {\it Calculation of generating function in many-body systems with quantum computers: Technical challenges and use in hybrid quantum classical methods},
arXiv:2104.08181.

\bibitem{Rui22} Edgar Andres Ruiz Guzman and Denis Lacroix, {\it Accessing ground-state and excited-state energies in a many-body system after symmetry restoration using quantum computers}, Phys. Rev. {\bf C 105}, 024324 (2022).

\bibitem{Rui23} Edgar Andres Ruiz Guzman and Denis Lacroix, {\it Restoring broken symmetries using quantum search “oracles”},  Phys. Rev. {\bf C 107}, 034310 (2023). 

\bibitem{Lac23}  D. Lacroix, E. A. Ruiz Guzman, and P. Siwach, {\it Symmetry breaking/symmetry preserving circuits and symmetry restoration on quantum computers}, Eur. Phys. J. {\bf A 59}, 3 (2023).


\bibitem{Fab21} Javier Faba, Vicente Mart\'in, and Luis Robledo, {\it Correlation energy and quantum correlations in a solvable model}, Phys. Rev. {\bf A 104}, 032428 (2021). 

\bibitem{Fab22} Javier Faba, Vicente Mart\'in, and Luis Robledo, {\it Analysis of quantum correlations within the ground state of a three-level Lipkin model},
Phys. Rev. {\bf A 105,} 062449 (2022).

\bibitem{Mom23} S. Momme Hengstenberg, Caroline E. P. Robin, Martin J. Savage, {\it Multi-Body Entanglement and Information Rearrangement in Nuclear Many-Body Systems}, Eur. Phys. J. {\bf A 59}, 231 (2023).


\bibitem{Rob21} Caroline Robin, Martin J. Savage, and Nathalie Pillet, {\it Entanglement rearrangement in self-consistent nuclear structure calculations}, Phys. Rev. {\bf C 103}, 034325 (2021).  

\bibitem{Joh23} Calvin W. Johnson, Oliver C. Gorton, {\it Proton-neutron entanglement in the nuclear shell model }, J. Phys. G: Nucl. Part. Phys. {\bf 50}, 045110 (2023). 

\bibitem{Per23}  A. P\'erez-Obiol, S. Masot-Llima, A. M. Romero, J. Men\'endez, A. Rios, A. Garc\'ia-S\`aez, B. 
Juli\'a-D\'iaz, {\it Quantum entanglement patterns in the structure of atomic nuclei within the nuclear shell model}, 
Eur. Phys. J. {\bf A 59}, 240 (2023). 

\bibitem{Gu23}  Chenyi Gu, Z. H. Sun, G. Hagen, T. Papenbrock, {\it Entanglement entropy of nuclear systems},
Phys. Rev. {\bf C 108}, 054309 (2023).

\bibitem{Rin80} P. Ring and P. Schuck, {\it The Nuclear Many-Body Problem} (Springer-Verlag, New-York, 1980).

\bibitem{Bla86} J. P. Blaizot and G. Ripka, {\it Quantum Theory of Finite Systems} (MIT Press, Cambridge, 1986).    
\bibitem{Ben03} M. Bender, P.-H. Heenen, and P.-G. Reinhard, {\it Self-consistent mean-field models for nuclear structure},  Rev. Mod. Phys. {\bf 75}, 121 (2003).



\bibitem{Rob18} L. M. Robledo, , T. R. Rodr\'iguez, and R. R. Rodr\'iguez-Guzm\'an, {\it Mean field and beyond description of nuclear structure with the Gogny force a review}, Journal of Physics G: Nuclear and Particle Physics {\bf 46},  013001 (2018).

\bibitem{She19}  J. A. Sheikh, J. Dobaczewski, P. Ring, L. M. Robledo, C. Yannouleas, {\it Symmetry restoration in mean-field approaches }, 
J. Phys. G: Nucl. Part. Phys {\bf 48}, 123001 (2021).



\bibitem{Zhe23} Muqing Zheng and Bo Peng and Nathan Wiebe and Ang Li and Xiu Yang and Karol Kowalski, {\it Quantum algorithms for generator coordinate methods}, Phys. Rev. Research {\bf 5}, 023200 (2023).



\bibitem{Hil53} D. L. Hill and J. A. Wheeler, {\it Nuclear constitution and the interpretation of fission phenomena}, Phys. Rev. {\bf 89}, 1102 (1953).


\bibitem{Lip65} H. J. Lipkin, N. Meshkov and A. J. Glick, {\it Validity of many-body approximation methods for a solvable model: (I) Exact solutions and 
perturbation theory},  Nucl. Phys. {\bf A 62}, 188 (1965).

\bibitem{Mes65} N. Meshkov, A. Glick, and H. Lipkin, {\it Validity of many-body approximation methods for a solvable model: (II). Linearization procedures},
Nucl. Phys. {\bf 62}, 199 (1965).

\bibitem{Gli65} A. Glick, H. Lipkin, and N. Meshkov, {\it Validity of many-body approximation methods for a solvable model: (III). Diagram summations}, 
Nucl. Phys. {\bf 62}, 211 (1965).


\bibitem{Hig19} Oscar Higgott and Daochen Wang and Stephen Brierley, {\it Variational Quantum Computation of Excited States}, Quantum {\bf 3}, p 156, (2019)



\bibitem{Sek22}  Kazuhiro Seki, Seiji Yunoki, {\it Spatial, spin, and charge symmetry projections for a Fermi-Hubbard model on a quantum computer}, Phys. Rev. {\bf A 105}, 032419 (2022). 


\bibitem{Tsu20}  T. Tsuchimochi, Y. Mori, and S. L. Ten-no, {\it Spin-projection for quantum computation: A low-depth approach to strong correlation}, Phys. Rev. Research {\bf 2}, 043142 (2020).

\bibitem{Tsu22} Takashi Tsuchimochi, Masaki Taii, Taisei Nishimaki, Seiichiro L. Ten-no, {\it Adaptive construction of shallower quantum circuits with quantum spin projection for fermionic systems}, Phys. Rev. Research {\bf 4}, 033100 (2022).

\bibitem{Rui23b} Edgar Andres Ruiz Guzman and Denis Lacroix, {\it Restoring symmetries in quantum computing using Classical Shadows}, arXiv:2311.04571. 


\bibitem{Bal69} R. Balian and E. Br\'ezin, {\it Nonunitary Bogoliubov transformations and extension of Wick's theorem}, 
Nuovo Cimento {\bf 64}, 37 (1969).

\bibitem{Per14} A. Peruzzo et al., {\it A variational eigenvalue solver on a photonic quantum processor}, 
Nat. Commun. {\bf 5}, 4213 (2014). 


\bibitem{Aga66} D. Agassi , H. J. Lipkin and N. Meshkov, {\it Validity of many-body approximation methods for a solvable model: (IV). The deformed Hartree-Fock solution}, Nucl. Phys. {\bf 86}, 321 (1966).
Nucl. Phys. {\bf 86} (1966) 321. 

\bibitem{Aga68} D. Agassi, {\it Validity of the BCS and RPA approximations in the pairing-plus-monopole solvable model}, Nucl. Phys. {\bf A 116}, 49 (1968).


\bibitem{Jor28} P. Jordan and E. Wigner, {\it \"Uber das Paulische \"Aquivalenzverbot},  Z. Phys. {\bf 47}, 631 (1928).

\bibitem{Lie61} Elliott Lieb, Theodore Schultz, Daniel Mattis, {\it Two soluble models of an antiferromagnetic chain}, Ann. of Phys. {\bf 16},  407 (1961).


\bibitem{Hol73} G. Holzwarth, {\it Four approaches to the function of inertia in a solvable model}, Nucl. Phys. 
{\bf A 207}, 545 (1973).

\bibitem{Rob92} L. M. Robledo, {\it Characterization of octupole correlations in the Lipkin model} Phys. Rev. 
{\bf C 46}, 238 (1992).

\bibitem{Sev06} A. P. Severyukhin, M. Bender, and P.-H. Heenen, {\it Beyond mean field study of excited states: Analysis within the Lipkin model} Phys. Rev.  {\bf C 74}, 024311 (2006).


\bibitem{Zha90} W.-M. Zhang, D.H. Feng and R. Gilmore, {\it Coherent states: Theory and some applications}, Rev. Mod. Phys. {\bf 62},  867 (1990).



\bibitem{Nie00}{M. A. Nielsen and I. L. Chuang, {\it Quantum information
and quantum computation}, Cambridge University Press (2000).}


\bibitem{Aru20} F. Arute, K. Arya, R. Babbush, D. Bacon, J.C. Bardin, R. Barends, S. Boixo, M. Broughton, B.B. Buckley, D.A. Buell et al., {\it Hartree- Fock on a superconducting qubit quantum computer}, Science {\bf 369}, 1084 (2020). 

\bibitem{Dal18} P.-L. Dallaire-Demers, J. Romero, L. Veis, S. Sim, A. Aspuru- Guzik, {\it Low-depth circuit ans\"atz for preparing correlated fermionic states on a quantum computer}, Quant. Sci. Technol. {\bf 4}, 045005 (2018). 



\bibitem{Sta20} N. H. Stair, R. Huang, and F. A. Evangelista, {\it A multireference
quantum Krylov algorithm for strongly correlated electrons},
J. Chem. Theory Comput. {\bf 16}, 2236 (2020).

\bibitem{Bha21a} K. Bharti and T. Haug, {\it Iterative quantum-assisted eigensolver},
Phys. Rev. {\bf A 104}, L050401 (2021).

\bibitem{Bha21b}  K. Bharti and T. Haug, {\it Quantum-assisted simulator}, Phys. Rev.
{\bf A 104}, 042418 (2021).




\bibitem{Qis21} Qiskit Development Team,
{\it Qiskit: An Open-source Framework for Quantum Computing}, (2021). 


\bibitem{Byr95} Byrd, R. H., Lu, P., Nocedal, J., \& Zhu, C., {\it A limited memory algorithm for bound constrained optimization.}, SIAM Journal on Scientific Computing {\bf 16}, 1190 (1995). 

\bibitem{Zhu97} Ciyou Zhu, Richard H. Byrd, Peihuang Lu, and Jorge Nocedal. 1997. {\it Algorithm 778: L-BFGS-B: Fortran subroutines for large-scale bound-constrained optimization.} ACM Trans. Math. Softw. {\bf 23}, 550 (1997). 

\bibitem{Li17} Li, Ying and Benjamin, Simon C., 2017. {\it Efficient variational quantum simulator incorporating active error minimization}, Phys. Rev. {\bf X 7}, 021050 (2017).

\bibitem{Tem17} Temme, Kristan and Bravyi, Sergey and Gambetta, Jay M. 2017. {\it Error mitigation for short-depth quantum circuits}, Phys. Rev. Lett. {\bf 119}, 180509 (2017).

\end{thebibliography}
\end{document}